Citation impact of papers published from six prolific countries:

A national comparison based on InCites data

Lutz Bornmann* & Loet Leydesdorff$

* Max Planck Society, Administrative Headquarters, Hofgartenstr. 8, 80539 Munich, Germany. E-mail: bornmann@gv.mpg.de

$ Amsterdam School of Communication Research (ASCoR), University of Amsterdam, Kloveniersburgwal 48, 1012 CX Amsterdam, The Netherlands. E-mail: loet@leydesdorff.net

**Abstract**

Using the InCites tool of Thomson Reuters, this study compares normalized citation impact values calculated for China, Japan, France, Germany, United States, and the UK throughout the time period from 1981 to 2010. The citation impact values are normalized to four subject areas: natural sciences; engineering and technology; medical and health sciences; and agricultural sciences. The results show an increasing trend in citation impact values for France, the UK and especially for Germany across the last thirty years in all subject areas. The citation impact of papers from China is still at a relatively low level (mostly below the world average), but the country follows an increasing trend line. The USA exhibits a relatively stable pattern of high citation impact values across the years. With small impact differences between the publication years, the US trend is increasing in engineering and technology but decreasing in medical and health sciences as well as in agricultural sciences. Similar to the USA, Japan follows increasing as well as decreasing trends in different subject areas, but the variability across the years is small. In most of the years, papers from Japan perform below or approximately at the world average in each subject area.

# 1 Introduction

Studies focusing on the international comparative performances of countries or territories have a long tradition. Leydesdorff and Wagner (2009) ascertain that "the USA is still outperforming all other countries in terms of highly cited papers and citation/publication ratios, and it is more successful than the EU in coordinating its research efforts in strategic priority areas like nanotechnology" (p. 23). The Science and Engineering Indicators of the US National Science Board (2012) report developments in international and US science and technology based on a comprehensive data base. This report shows that the combined share of published articles of researchers in the EU and the USA "decreased steadily from 69% in 1995 to 58% in 2009. In little more than a decade, Asia's world article share expanded from 14% to 24%, driven by China's 16% average annual growth dominated world article production" (p. 0-10). With regard to citation impact, the analyses of the National Science Board (2012) reveal that "U.S. articles continue to have the highest citation rates across all broad fields of S&E" (p. 5-43). On the excellence level, the USA has published 76% more articles than expected among the top-1% most cited articles in 2010; scientists from the EU have published top-1% articles 7% less than expected.

Two recently published reports, although based on different data bases (Web of Science, WoS, Thomson Reuters, and Scopus, Elsevier), came to the same conclusion: In comparison with other prolific countries, the UK ranks top (first and second, respectively) in citations per paper (Marshall & Travis, 2011). According to Adams (2010) "UK performance is on a rising trajectory, whereas the USA research base has at best plateaued in performance and – on some estimates – is now in decline … the UK has now overtaken the US on average and … remains well ahead of Germany and France. Japan has dropped well behind its G7 comparators and its average impact is now similar in performance to China" (p. 8). Whereas Adams (2010) used WoS data, Elsevier (2011) reports the following results based on Scopus data: "The UK's field-weighted citation impact … is well above the world benchmark, but is slightly lower than that of the US, ranking it 2nd in the G8 and in the comparator group but with a growth rate of 1.1% compared to the US's -0.5% per year since 2006" (p. 28).

In this study, the citation impact analyses of Adams (2010) and Marshall and Travis (2011) are extended by (1) the consideration of a longer publication window (1981 to 2010 instead of 1991 to 2010), (2) a differentiation in four broad subject areas (instead of an analysis across all areas and some selected fields), and (3) the use of statistical procedures to investigate the citation impact trends across the years. Following Adams (2010) and Marshall and Travis (2011), the analyses of this study refer to the six most prolific countries: China, Japan, France, Germany, the United States, and the UK. Adams (2010) justifies the selection of these six countries as follows: they "now account for almost two-thirds of the global total of research publications indexed in Thomson Reuters Web of Knowledge$^{SM}$. France and Germany are the UK's major partners in the European Research Area. The USA and Japan are its key collaborators in other regions. China is the new dynamic factor in the global geography of science, as we have reported previously. Its rapid rise demonstrates a different model in research policy and demands a reanalysis of the certainties that underpinned the trans-Atlantic research axis of the last half-century" (p. 3).

This study is based on data from the relatively new InCites tool of Thomson Reuters, which facilitates national comparisons across long time periods using normalized citation impact values.

## 2 Methods

### 2.1 InCites

InCites (Thomson Reuters; http://incites.thomsonreuters.com/) is a web-based research evaluation tool allowing to assess the productivity and citation impact of institutions and countries. The global comparisons module provides citation metrics from the WoS for the evaluation of research output of institutions and countries. The metrics are generated from a dataset of 22 million WoS papers from 1981 to 2010. The metrics for country-specific comparisons are created based on address criteria using the whole-counting method (i.e. addresses of authors having published the papers). Thomson Reuters uses whole counting: Counts are not weighted by numbers of authors or number of addresses.

Country-specific metrics can be downloaded as a national comparison report in Excel format. As subject area scheme for this study, the main categories of the Organisation for Economic Co-operation and Development (2007; OECD) were used. (InCites provides six further schemes, e.g. the 22 subject areas provided by Thomson Reuters in the Essential Science Indicators. A concordance table between the OECD categories and the WoS subject categories is also provided.) In contrast to the other schemes, the OECD scheme enables the use of six broad subject categories for WoS data: (1) natural sciences, (2) engineering and technology, (3) medical and health sciences, (4) agricultural sciences, (5) social sciences, and (6) humanities. Each subject category incorporates subordinate fields. For example, agricultural sciences include the following subordinate subject fields: (1) agriculture, forestry, and fisheries; (2) animal and dairy science; (3) veterinary science; (4) agricultural biotechnology; and (5) other agricultural sciences.

For each broad subject category, the countries data (InCites$^{TM}$ Thomson Reuters, 2011) was downloaded as an Excel sheet and imported in Stata (StataCorp., 2011) for the statistical analysis. However, the numbers for the social sciences and humanities were not included in the analysis. According to Blockmans and Thomassen (2005) "one can scarcely expect researchers in the humanities and the social sciences to accept the performance indicators used in the natural sciences as valid in their own field. As a result, alternative methods are required. The success of any evaluation procedures and instruments depends on their being accepted by the relevant researchers" (p. 5).

**2.2  Citation metrics and statistical procedures**

To have reliable citation impact values in this study, only those publication years are included in the statistical analysis for a subject area in which at least 100 papers are available. For example, data for China in agricultural sciences were not included for the years 1981 to 1989. Following the presentation of Marshall and Travis (2011) and Adams (2010), we show in the following normalized citation impact values for the six different countries. Thomson Reuters calculates the mean citation rate to a country's set of publications and then divide it by the mean of all publications in that subject area. A value of 1 for a specific country in a specific subject area indicates that the citation impact of papers published by scientists in this country is no more and no less than the average impact of papers in this subject area.[1] If this value adds up to, e.g., 1.2 the corresponding papers were cited on average 20 percentage points above the average in this area.

For each country in a specific subject area analysis, the Spearman's rank-order coefficient ($\rho$) for the correlation between publication year and citation impact is calculated (Sheskin, 2007). The coefficients support the interpretation of the trend results beyond the visual inspection of the curves. A statistically significant coefficient ($p<.01$, two-sided tested; the critical absolute value for the coefficient is .478) (see here Glass & Stanley, 1970) points out whether the "true" coefficient (the population coefficient) is likely to be above 0. Since the coefficients are calculated on the basis of rank data, they do not consider the size of impact differences between two publication years but only the existence of a difference. This means that a country with a trend of increasing impact values may yet have a smaller absolute increase between 1981 and 2010 than another country with a (more or less) random distribution of values but with some specific impact boosts during the time period.

---

[1] Because normalization is pursued for a single category in each case, the denominator is a constant and therefore the issue of the difference between "rates of averages" or "averages of rates" does not play a role in this normalization (Egghe, 2012; Gingras & Larivière, 2011).

To measure the variability of citation impact values across the publication years, the standard deviation (SD) is calculated for each country and each field category. The standard deviation indicates the extent of deviations from a country's mean citation impact across all publication years. A relatively small standard deviation, e.g., indicates that the impact values do not deviate from the mean across all years to a large extent.

## 3 Results

Figure 1 shows, for six countries, citation impact values calculated relatively to the four subject areas. Two points should be considered in the interpretation of the results: (1) Larger citation impact differences between two following years for one country can rather be the effect of lower paper numbers than of significant performance differences. Since authors from the USA publish more papers than all other countries, the variation is accordingly low. (2) The longer the citation window, the more reliable is the performance estimation for a paper (Research Evaluation and Policy Project, 2005).Therefore, the most recent publication years in Figure 1 should be interpreted with due care (see, e.g., the outliers for 2010 in the citation impact values in the case of agricultural sciences).

If we compare the results for the different subject areas in Figure 1 with the figures published by Marshall and Travis (2011) and Adams (2010) across all subject areas, the correspondence is largest for natural sciences. In this subject area we see a similar pattern with a sharp increase in the impact of papers from the UK during the last four years. The UK is the only country which surpasses the performance of the USA across the last thirty years. It is interesting to see for natural sciences that all countries with the exception of the USA have very high correlation coefficients (China: $\rho=.89$, UK: $\rho=.84$, Germany: $\rho=.98$, France: $\rho=.98$, Japan: $\rho=.77$) pointing out increasing trends of citation impact values. This trend is especially visible for the European countries (France, Germany, and the UK). On a low performance level, the trend is also visible for China but it has been declining for the last two years. Studies in the next few years will show whether this trend will continue or not.

For engineering and technology, a different pattern is visible in Figure 1 than for natural sciences. Not only Germany (ρ=.96) and China (ρ=.65), but also the USA (ρ=.7) show an increasing citation impact trend on a relatively high level (these three countries have the highest correlation coefficients in the graph). The reverse trend is visible for Japan (ρ=-.49, but this coefficient is statistically not significant). In the next graph in the figure, medical and health sciences show clearly increasing trends for Germany (ρ=1) and France (ρ=1) but a decreasing trend for the USA (ρ=-.64). However, whereas the decreasing trend of the USA occurs on a high impact level across the past thirty years, Germany and France improved their impact from a very low performance level in the 1980s to a very high level (25% above the subject area average) for the last two years. Agricultural sciences – as the last subject area considered in Figure 1 – are characterized by increasing trends of China (ρ=.66), France (ρ=.91), and Germany (ρ=.99), but a decreasing trend of the USA (ρ=-.87). Across the whole time period, the UK performs on a very high level of at least 25% above the subject area average.

Table 1 shows the minima, maxima, means, and standard deviations of citation impact values for the different subject areas and countries. Although the USA has in most of the subject areas in Figure 1 a statistically significant trend (increasing or decreasing), the standard deviations are in all subject areas very small (SD<.1). This means that there is only a low level of variability between the publication years. A different situation is indicated for Germany (SD is between .15 and .25) and China (SD is between .13 and .16). These countries have a high level of variability and a dynamically increasing trend (as Figure 1 points out).

# 4     Discussion

In broad agreement with other studies (described in the Introduction section), this study shows an increasing trend of normalized citation impact values for France, the UK and especially for Germany across the last thirty years in four subject areas. For France and Germany, this may also be an internationalization effect. The scientists of both countries have published increasingly in English. Thus, one measures not only scientific performance, but also a shift of performance. The citation impact of papers from China is still on a relatively low level (mostly below the world average), but the country follows an increasing trend line. The USA shows a relatively stable pattern of high citation impact values across the years. With small impact differences between the publication years, the US trend is increasing in engineering and technology but decreasing in medical and health sciences as well as in agricultural sciences. Similar to the USA, Japan follows increasing as well as decreasing trends in different subject areas, but the variability across the years is generally small. In most of the years, papers from Japan perform below or approximately at the world average in a subject area.

As a first limitation of the study, the use of arithmetic average rates of citations should be mentioned. There are dangers in focusing on the measures of central tendency: in the face of non-normally distributed citation data, the arithmetic mean value can give a distorted picture of the kind of distribution (Bornmann, Mutz, Neuhaus, & Daniel, 2008), "and it is a rather crude statistic" (Adler, Ewing, Taylor, & Hall, 2009, p. 1). As the distribution of citation frequencies is usually right skewed, distributed according to a power law, arithmetic average rates of citations show mainly where papers with high citation rates are to be found (Bornmann, Mutz, Marx, Schier, & Daniel, 2011).

Particularly in bibliometric analysis the use of percentile rank scores for evaluative purposes is very advantageous, as no assumptions have to be made about the distribution of citations (Leydesdorff, Bornmann, Mutz, & Opthof, 2011). Although some national comparison studies additionally report results based on percentiles (National Science Board, 2012), the most commonly used approach is a presentation of (normalized) average citation rates. Since InCites does not provide percentiles in the global comparison module, this study followed this approach. For future national comparisons, studies based on percentiles are very desirable (see here, e.g., the new Leiden ranking at http://www.leidenranking.com/ for the comparison of institutions) (Bornmann, de Moya Anegón, & Leydesdorff, 2012).

A second limitation of our study concerns the use of four broad fields to normalize the results. Performing normalization at the level of these broad fields means that a high citation impact of a country in a particular field may be a consequence of the fact that most of the activity of the country in this field takes place in subfields with a relatively high citation density. So differences in citation impact may merely reflect differences in the research activities of countries (rather than differences in countries' actual scientific impact). Also, an increase (or decrease) of a country's citation impact in a particular field may also be a consequence of a shift of the country's research activities from subfields with a lower (higher) citation density to subfields with a higher (lower) citation density.

## 5 Conclusion

With the global comparisons module, InCites offers a unique opportunity to study the normalized citation impact of countries. Although the approach has its limitations, (1) the consideration of a long publication window (1981 to 2010), (2) a differentiation in (broad) subject areas, and (3) the use of statistical procedures allow for an insightful investigation of national citation impact trends across the years.

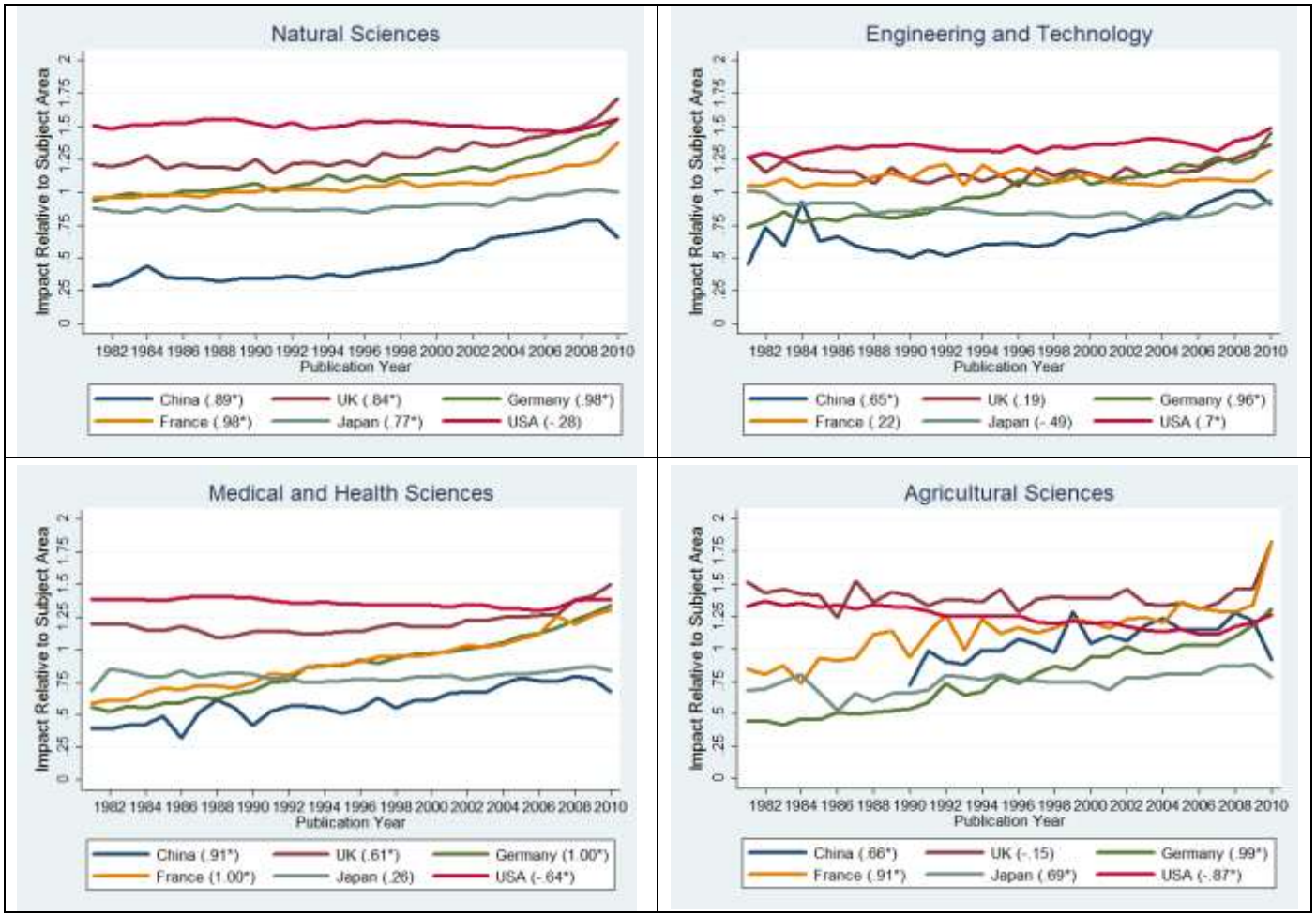

Figure 1. Citation impact of six countries (China, UK, Germany, France, Japan, and USA) calculated relatively to four subject areas (natural sciences; engineering and technology; medical and health sciences; and agricultural sciences). For each country, the Spearman's rank-order coefficient for the correlation between publication year and citation impact is given; an asterisk signifies a statistically significant coefficient (the "true" coefficient, the population coefficient, is likely to be above 0). A high correlation coefficient indicates an increasing or decreasing trend in citation impact values. Source: InCites[TM] Thomson Reuters (2011).

Table 1. Minimum, maximum, mean, and standard deviation of citation impact values by subject area and country (N is the number of publication years included in the analysis)

|  | **China** | **UK** | **Germany** | **France** | **Japan** | **USA** |
|---|---|---|---|---|---|---|
| **Natural Sciences** | | | | | | |
| *N* | 30 | 30 | 30 | 30 | 30 | 30 |
| Minimum | 0.29 | 1.14 | 0.93 | 0.95 | 0.85 | 1.45 |
| Maximum | 0.78 | 1.71 | 1.55 | 1.38 | 1.01 | 1.55 |
| Mean | 0.47 | 1.30 | 1.13 | 1.06 | 0.90 | 1.51 |
| Standard deviation | 0.16 | 0.13 | 0.15 | 0.10 | 0.05 | 0.03 |
| **Engineering and Technology** | | | | | | |
| *N* | 30 | 30 | 30 | 30 | 30 | 30 |
| Minimum | 0.45 | 1.05 | 0.73 | 1.03 | 0.77 | 1.25 |
| Maximum | 1.01 | 1.35 | 1.45 | 1.21 | 1.01 | 1.48 |
| Mean | 0.69 | 1.16 | 1.00 | 1.10 | 0.87 | 1.34 |
| Standard deviation | 0.15 | 0.07 | 0.19 | 0.05 | 0.05 | 0.05 |
| | | | | | | |
| **Medical and Health Sciences** | | | | | | |
| N | 30 | 30 | 30 | 30 | 30 | 30 |
| Minimum | 0.33 | 1.09 | 0.53 | 0.59 | 0.69 | 1.30 |
| Maximum | 0.79 | 1.50 | 1.34 | 1.30 | 0.87 | 1.41 |
| Mean | 0.58 | 1.20 | 0.87 | 0.90 | 0.80 | 1.36 |
| Standard deviation | 0.13 | 0.09 | 0.24 | 0.20 | 0.04 | 0.03 |
| | | | | | | |
| **Agricultural Sciences** | | | | | | |
| N | 21 | 30 | 30 | 30 | 30 | 30 |
| Minimum | 0.72 | 1.24 | 0.42 | 0.73 | 0.53 | 1.11 |
| Maximum | 1.28 | 1.83 | 1.30 | 1.83 | 0.88 | 1.36 |
| Mean | 1.06 | 1.41 | 0.76 | 1.13 | 0.74 | 1.24 |
| Standard deviation | 0.14 | 0.10 | 0.25 | 0.21 | 0.08 | 0.08 |